\documentclass[namedreferences]{SolarPhysics}

\usepackage[optionalrh,solaenum]{spr-sola-addons} % For Solar Physics 
\usepackage{graphicx}        % For eps figures, newer & more powerfull
\usepackage{color}           % For color text: \color command

\usepackage[hyphens]{url}            % For breaking URLs easily through lines

\usepackage{solaheader}
\newcommand{\solareceiveddate}{17 May 2013} % Remove the leading % and fill in the values when known; solaheader must be used
\newcommand{\solaaccepteddate}{5 August 2013}

\ifx \doiurl    \undefined \def \doiurl#1{\href{http://dx.doi.org/#1}{\textsf{#1}}}\fi
\ifx \adsurl    \undefined \def \adsurl#1{\href{http://adsabs.harvard.edu/abs/#1}{\textsf{#1}}}\fi
\ifx \arxivurl  \undefined \def \arxivurl#1{\href{http://arxiv.org/abs/#1}{\textsf{#1}}}\fi
\ifx \urlurl    \undefined \def \urlurl#1{\href{http://#1}{\textsf{#1}}}\fi
\ifx \mailtourl    \undefined \def \mailtourl#1{\href{mailto:#1}{\textsf{#1}}}\fi

\def\arcdeg{\hbox{$^\circ$}}
\def\arcmin{\hbox{$^{\prime}$}}
\def\arcsec{\hbox{$^{\prime\prime}$}}
\def\gtrsim{\mathrel{\hbox{\rlap{\hbox{\lower4pt\hbox{$\sim$}}}\hbox{$>$}}}}
\def\sun{\hbox{$\odot$}}

\newcommand{\adv}{    {\it Adv. Space Res.}}

\newcommand{\aap}{    {\it Astron. Astrophys.}}

\newcommand{\apj}{    {\it Astrophys. J.}}
\newcommand{\apjl}{    {\it Astrophys. J. Lett.}}

\newcommand{\jgr}{    {\it J. Geophys. Res.}}

\newcommand{\solphys}{{\it Solar Phys.}}

\newcommand{\ssr}{    {\it Space Sci. Rev.}}

\newcommand{\apjs}{    {\it Astrophys. J. Supp.}}
\newcommand{\lrsp}{    {\it Living Rev. Solar Phys.}}

\begin{document}

\begin{article}

\begin{opening}

\title{The Association of Solar Flares with Coronal Mass Ejections
  During the Extended Solar Minimum }

\author{N.V.~\surname{Nitta}$^{1}$\sep
        M.J.~\surname{Aschwanden}$^{1}$\sep
        S.L~\surname{Freeland}$^{1}$\sep
        J.R.~\surname{Lemen}$^{1}$\sep
        J.-P.~\surname{W\"{u}lser}$^{1}$\sep
        D.M.~\surname{Zarro}$^{2}$
}
\runningauthor{N.V. Nitta {\it et al.}}
\runningtitle{Flares and CMEs during the Minimum}

   \institute{$^{1}$ Lockheed Martin Solar and Astrophysics
     Laboratory, A021S, Building 252, 3251 Hanover Street, Palo Alto, CA
     94304 USA
                     email: \mailtourl{nitta@lmsal.com} email:
                     \mailtourl{aschwanden@lmsal.com} email:
                     \mailtourl{freeland@lmsal.com} email:
                     \mailtourl{lemen@lmsal.com} email:
                     \mailtourl{wuelser@lmsal.com} \\
$^{2}$ ADNET, Systems, Inc., and NASA Goddard Space Flight Center, Greenbelt
     20771 USA   email: \mailtourl{dominic.m.zarro@nasa.gov}
             }

\begin{abstract}
We study the association of solar flares with coronal mass
ejections (CMEs) during the deep, extended solar minimum of 2007\,--\,2009,
using extreme-ultraviolet (EUV) and white-light (coronagraph) images 
from the {\it Solar Terrestrial Relations Observatory} (STEREO). 
Although all of the fast (v $>$ 900~km~s$^{-1}$) {\it and} wide ($\theta
>$ 100$\arcdeg$) CMEs are associated with a flare that is at least
identified in GOES soft X-ray light curves,   
a majority of flares with relatively high X-ray intensity 
for the deep solar minimum 
({\it e.g.} $\gtrsim$1$\times$10$^{-6}$~W m$^{-2}$ or C1)
are not associated with CMEs. 
Intense flares tend to occur in active regions with strong
and complex photospheric magnetic field, 
but the active regions that produce CME-associated flares tend to be small,
including those that have no sunspots and therefore no 
NOAA active-region numbers. 
Other factors on scales comparable to and larger
  than active regions seem to exist that contribute to the association
  of flares with CMEs.
% This suggests that large-scale field may be
% important for a flare to be fully eruptive. 
We find the possible low coronal signatures of CMEs, namely eruptions, 
dimmings, EUV waves, and Type III bursts, in 
91\,\%, 74\,\%, 57\,\%, and 74\,\%, respectively, of the 35 flares
that we associate with CMEs.  None of these observables can fully
replace direct observations of CMEs by coronagraphs.

\end{abstract}
\end{opening}
%-------------------------------------------------

\section{Introduction}
     \label{S-Introduction} 

The origin of coronal mass ejections (CMEs) is still not well
understood to the level that we can predict them in advance, although
considerable progress has been made in recent years in observations
and numerical models (see, {\it e.g.}, \citeauthor{PFChen11},
\citeyear{PFChen11}, and references therein.)  
One of the associated phenomena is the solar
flare.  The solar flare is characterized by a sudden increase in
electromagnetic radiation, as a result of heating and particle
acceleration, which are usually thought to result from 
magnetic reconnection.
One problem appears to be the lack of consensus as to the
role of magnetic reconnection in CMEs, which eject 
coronal magnetized plasma into the heliosphere.

Intense flares are often observed around the times of
extremely energetic CMEs, such as several of those during the
October\,--\,November 2003 period 
({\it e.g.} \citeauthor{Gopalswamy05}, \citeyear{Gopalswamy05}).
However, less than half of CMEs are associated with
flares \cite{Munro79}.  Moreover, it has been reported that certain CMEs, 
typically during solar minimum, may leave no
observable signatures in the corona \cite{Robbrecht09b, Ma10}, not to
mention flares.
Nevertheless, a flare-associated CME tends to show a
characteristic kinematic pattern: that is, starting fast and then
decelerating \cite{MacQueen83, Sheeley99}.

The association of flares with CMEs 
increases with their peak soft X-ray flux.
By definition, a CME is observed with white-light
coronagraphs.  Their detection sensitivity is higher when the source region is
located near the limb because of the way that Thomson scattering works
\cite{Yashiro05}.
According to \citeauthor{Yashiro05} \shortcite{Yashiro05}, 
who carefully made associated CME-flare pairs during 1996\,--\,2001, 
the location-averaged CME association rate
is roughly 20\,\%, 50\,\% and 90\,\% for C-class, M-class, and X-class flares,
respectively.  When the two phenomena are associated,
    they tend to be closely related both temporally 
({\it e.g.} \citeauthor{Zhang01}, \citeyear{Zhang01};
\citeauthor{Temmer10}, \citeyear{Temmer10}) and spatially 
({\it e.g.} \citeauthor{Yashiro08}, \citeyear{Yashiro08}).

In this article, we study the association of flares with CMEs during
the extended minimum after Solar Cycle 23, primarily using 
remote-sensing data
from the {\it Solar Terrestrial Relations Observatory} 
(STEREO: \citeauthor{Kaiser08}, \citeyear{Kaiser08}).  
Flares were observed by the {\it Extreme-Ultraviolet Imager} (EUVI:
\citeauthor{Wuelser04}, \citeyear{Wuelser04}; \citeauthor{Howard08},
\citeyear{Howard08}).  To associate flares with CMEs, 
we compare EUVI with white-light data from the 
{\it Large Angle Spectroscopic Coronagraph} (LASCO:
\citeauthor{Brueckner95}, \citeyear{Brueckner95}) on 
the {\it Solar and Heliospheric Observatory} (SOHO), and the COR-1 and COR-2
coronagraphs \cite {Howard08} on STEREO.
The ambiguity in associating flares with CMEs has been considerably reduced,
thanks to the availability of high cadence COR-1
data that cover the corona between 1.4\,R$_{\sun}$ and 4.0\,R$_{\sun}$
in heliocentric distance, and also to the solar minimum conditions in which successive
flares or CMEs tend to occur farther apart 
in both 
time and space than
when the solar activity is high.
In the next section, we briefly review the solar minimum in question, 
and active regions and flares during the period.
After describing the data used in this study, we present the analysis in
Section~3.  
The results are given in
Section~4 with some discussion of selected examples.  
In Section~5 we summarize our study.

\section{Flares and Active Regions During 2007\,--\,2009} %%%%%%%%%%%%%%%%%%%%%%%%%%%%%%%%%%%%%%%%
      \label{General}      

\begin{figure}    %%%%%%%%%%%%%%%%%% FIGURE 1 
 \centerline{\includegraphics[width=1.00\textwidth,clip=]{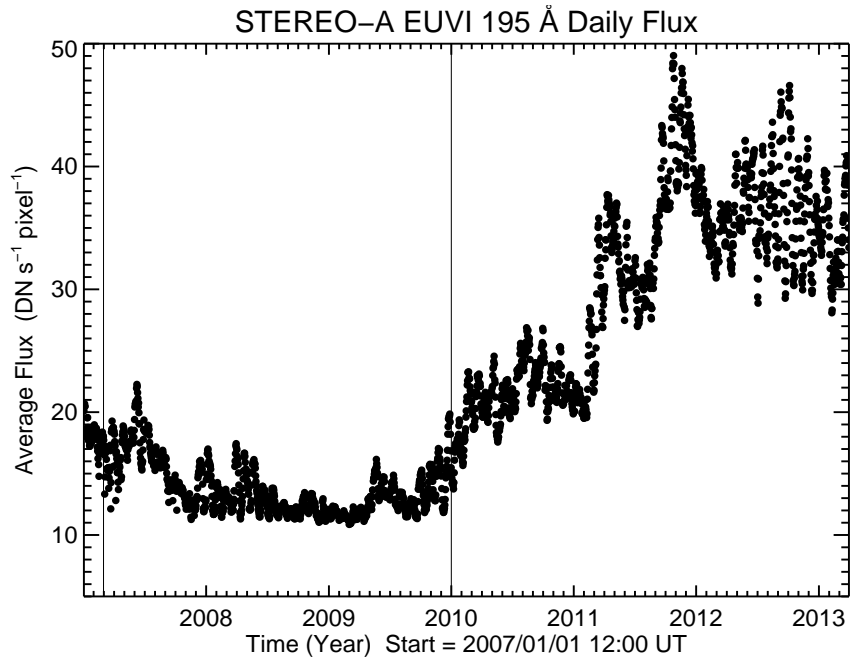}}
% {./euvi_195_long_term_a_only4paper_rev.eps}}
              \caption{Daily EUV flux as obtained by averaging EUVI
                195~\AA\ images taken around 12 UT.  The vertical
                lines indicate the interval of interest of this work.
}
   \label{euvi_long_term}
\end{figure}

Following the last X-class flare on 14 December 2006 in Solar Cycle
23, which occurred before the imagers on STEREO started to be
operational, not many eruptive events
with intense X-ray emission were observed, 
as we show in Section~4. This situation apparently changed in
February 2010, when several CMEs were observed in association with
GOES M-class flares.  
In order to give a glimpse of the last solar minimum,
we show in Figure~1 the daily EUV flux from EUVI 195~\AA\ on STEREO-Ahead (A).  The average
flux per pixel was processed using images taken around 12 UT.  This
plot also shows that, around February 2010, the spatially-averaged
EUV flux resumed the level as of the beginning of 2007.

A catalog of flares observed by EUVI was 
already published by \citeauthor{Aschwanden09} \shortcite{Aschwanden09}.  
It contains a total of 185 flares up to 4 November 2008 
that were either above the GOES C1 level 
(1$\times$10$^{-6}$~W m$^{-2}$ in the 1\,--\,8~\AA\ band) or
detected by the {\it Reuven Ramaty
High Energy Solar Spectroscopic Imager} (RHESSI: \citeauthor{Lin02},
\citeyear{Lin02}).  Its emphasis was on the properties of
flares as observed in the EUV wavelengths.   However, the CME-flare
association was not discussed except in a handful of rare events.
Although \citeauthor{Aschwanden09} \shortcite{Aschwanden09} found a 
report of CME occurrence from LASCO, COR1, or COR2 
around the times of about one third of
the flares, such a report does not
necessarily mean that the flare is associated with the CME, 
unless we make sure that 
the two phenomena occur close both in space and time.
Here we focus on
the association of flares and CMEs during the extended minimum 
after Solar Cycle 23,
by comparing EUV and coronagraph images.  As shown in Section 4,
we find that a much smaller number of flares were associated
with CMEs than were indicated by 
\citeauthor{Aschwanden09} \shortcite{Aschwanden09}.

During March 2007\,--\,December 2009, 11 M-class and 95 C-class flares
occurred.  There were no X-class flares.  As in 
\citeauthor{Aschwanden09} \shortcite{Aschwanden09}, 
we study all of these M-class and C-class flares, and also some
B-class and A-class flares.  However, inclusion of the latter flares
depends on whether they are associated either with a CME or
CME-related signatures, rather than on their detection by RHESSI.

\begin{table}
\caption{Active regions that produced C-class and M-class flares
  during March 2007\,--\,December 2009 
}
\label{T-simple}
\tabcolsep 3.8pt
\begin{tabular}{ccrrrrrrrr}     % define the column alignment
                           % l: left, c: center, r: right
  \hline                   % horizontal line
1  & 2 & 3  & 4 & 5 & 6 & 7 & 8 & 9 & 10 \\
AR & Date & Lat. & Long. & Pol.  & Area & $\log R$ &
Comp. & M & C \\
\hline
10953 &  1 May 2007 & -11  & 308 & O & 520 & 4.0 & $\beta \gamma \delta$    & 0  & 2  \\
10956 & 19 May 2007 & 3    & 71  & ? & 300 & 4.4 & $\beta \gamma \delta$    & 0  & 2  \\
10960 &  7 Jun 2007 & -7   & 178 & O & 540 & 4.3 & $\beta \gamma \delta$    & 10 & 17 \\
10962 &  3 Jul 2007 & -9   & 189 & O &  60 & 2.5 & $\beta$                  & 0  & 2  \\
10963 & 15 Jul 2007 & -6   &  57 & O & 530 & 3.6 & $\beta \gamma$           & 0  & 19 \\
10966 &  9 Aug 2007 & -6   &  66 & O &  40 & 2.9 & $\beta$                  & 0  &  2 \\
10969 & 27 Aug 2007 & -7   & 188 & O & 540 & 2.9 & $\beta$                  & 0  &  1 \\
10978 & 11 Dec 2007 & -8   & 223 & O & 340 & 4.3 & $\beta \gamma \delta$    & 0  & 10 \\
10980 &  8 Jan 2008 & -7   & 237 & O &  30 & 3.0 & $\beta$                  & 0  &  5 \\
10988 & 30 Mar 2008 & -8   & 237 & O & 300 & 3.3 & $\beta$                  & 0  &  1 \\
10989 &  1 Apr 2008 & -12  & 206 & O &  80 & 2.8 & $\beta$                  & 1  &  0 \\
11007 &  1 Nov 2008 &  35  & 254 & N &  80 & 3.8 & $\beta$                  & 0  &  2 \\
N-AR$^{*}$  & 4 Dec 2008 & -25  & 170 & ? &  ?  & ?   & ?                        & 0  &  1 \\
11024 & 5 Jul 2009 & -27  & 247 & N & 230 & 3.7 & $\beta$                  & 0  &  2 \\
11026 & 27 Sep 2009 & -31  & 220 & N &  70 & $<$2.0 & $\beta$                  & 0  &  1 \\
11029 & 25 Oct 2009 &  15  & 212 & N & 380 & 4.2 & $\beta \gamma$           & 0  & 11 \\
11034 & 15 Dec 2009 &  19  & 252 & N &  20 & 4.0 & $\beta$                  & 0  &  1 \\
11035 & 15 Dec 2009 &  30  & 249 & N & 310 & 4.4 & $\beta \delta$           & 0  &  6 \\
11036 & 18 Dec 2009 & -28  & 209 & N &  70 & 2.9 & $\beta$                  & 0  &  3 \\
11038 & 19 Dec 2009 &  16  & 204 & N &  20 & $<$2.0 & $\beta$                  & 0  &  2 \\
11039 & 31 Dec 2009 & -27  &  53 & N & 220 & 3.9 & $\beta$                  & 0  &  5 \\

\hline
\end{tabular}

\noindent
1: NOAA AR number. 2: Date of central meridian (CM) passage.  
3: Latitude. 4: Carrington longitude. 
5: Whether the AR belongs to the old (O) or new (N)
solar cycle (23 or 24), depending on how the
positive and negative polarities are aligned East\,--\,West 
assuming Hale's law.  6: Maximum sunspot area 
(in microhemispheres) during the disk passage. 7: Maximum $R$ parameter 
\cite{Schrijver07} in logarithmic during the disk passage. 
8: Maximum magnetic complexity in the Mt. Wilson scheme during the disk passage.  
9: Number of M-class flares. 10: Number of C-class flares. \\

\noindent
$^{*}$ Not a NOAA active region.

\end{table}

In Table 1, we show the basic properties of 
active regions (ARs) in which M- and
C-class flares occurred in the period.
Only two ARs were responsible for the M-class flares 
and a total of 20 ARs produced C-class flares.  
No data analysis is needed for these properties 
but the $R$ parameter \cite{Schrijver07}, 
which is the total unsigned magnetic flux near the polarity inversion lines, 
and is used as a measure of magnetic
free energy in an AR. 
According to \citeauthor{Schrijver07} \shortcite{Schrijver07}, 
an AR with $R$ can, within 24 hours,
produce flares whose peak GOES flux is up to $1.2 \times 10^{-8} R$
W\,m$^{-2}$, which is X1.2 for $\log R = 4.0$.   We compute $R$ for
every magnetogram taken by the {\it Michelson Doppler Imager} 
(MDI: \citeauthor{Scherrer95}, \citeyear{Scherrer95}) onboard SOHO 
that contains the AR within
0.7\,R\,$_{\sun}$ from disk center, and put the maximum value of $\log R$ in
the seventh column.   

Based on the latitude and Carrington longitude, only AR 10960 and AR 10978 
survived more than one solar rotation.  They were not traceable in the
third rotation, however.  
This is different from the solar minimum around 1996, in which
a single region, which was AR~7978 in the first rotation, 
survived for four months 
({\it e.g.} \citeauthor{vanDriel98}, \citeyear{vanDriel98}).

Active regions started to follow the Cycle 24 polarity sometime in 2008, and 
AR~11024 (July 2009) was the first region with this polarity whose sunspot area 
exceeded 200 microhemispheres.
Note that the polarity of AR~10956 is not given in Table~1.  This is because
it contained a second bipole that violated Hale's law
\cite{Bone09}.  The presence of such a bipole and its interaction with
the primary bipole, which followed Hale's law, 
may have made the AR more CME-productive
(see Section 4), although the most intense flare in the AR was only C2.9.
The polarity is unknown also for the region that produced a C-class
flare in December 2008.  The region had no sunspot and therefore was
not a NOAA AR.

\section{Associating Flares with CMEs} %%%%%%%%%%%%%%%%%%%%%%%%%%%%%%%%%%%%%%%%
      \label{Procedure} 

\subsection{Data}

We limit our analysis to flares during March 2007\,--\,December 2009. This is 
partly to
% expolore the peculiarities of the extended solar minimum well after
% the last notable activity in Solar Cycle 23 in terms of
% the CME-flare relation.  
isolate the solar minimum away from the notable activities
in Solar Cycles 23 and 24.
Each flare detected by the GOES {\it X-ray Sensor} (XRS)
is studied using EUVI data.  EUVI is a normal-incidence
telescope on board STEREO with four EUV wavelength bands 
(171~\AA, 195~\AA, 284~\AA\ and 304~\AA) similar to 
those of the {\it Extreme-ultraviolet Imaging Telescope} 
(EIT: \citeauthor{Boudin95}, \citeyear{Boudin95}) on SOHO.
EUVI observes the Sun in 2048$\times$2048 1.6$\arcsec$ pixels. The 
field of view (FOV) in the North\,--\,South and East\,--\,West 
directions extends to $\approx$1.7\,R$_{\sun}$.

EUVI data are analyzed primarily to study three signatures closely related to
CMEs ({\it e.g.} \citeauthor{Hudson01}, \citeyear{Hudson01}), namely
eruptions, coronal dimming and coronal waves.  
Eruptions in EUV may be
% can be observed in all the four wavelength bands of EUVI.  They are
a lower-temperature counterpart of X-ray ejections that are often
associated with CMEs \cite{Nitta99,Kim05,Tomczak12}.
% Based on SOHO EIT observations, 
Dimming and waves
are usually best captured in images in the 195~\AA\ band 
({\it e.g.} \citeauthor{Wills_Davey99}, \citeyear{Wills_Davey99}), 
dominated by Fe~{\sc xii} lines, 
even though significant contributions from the Fe~{\sc xxiv} line
due to hot flare plasma are seen during intense flares \cite{Nitta13a}.

Despite the importance of the 195~\AA\ band, 
the typical cadence of images in this band was only ten minutes because 
171~\AA\ images were given higher priority.  This was changed in
August 2009, when the 195~\AA\ band was finally made the primary
wavelength band.
The typical cadences before the change were 2.5,
20, and 10 minutes for the 171~\AA, 284~\AA, and 304~\AA\ bands,
respectively.
After August 2009, in addition to 195~\AA\ images at the five\,--\,minute
cadence, 304~\AA\ images have been taken every ten minutes.
Images in the remaining two bands have been taken at lower cadences 
except during occasional campaign observations.
Early in the mission the same flares were observed by both STEREO
spacecraft.  But more flares were observed by only one spacecraft at
later times due to increased separation of STEREO from the Sun--Earth line. 
With the rate of
$\approx$22$\arcdeg$/year drift of each STEREO
spacecraft,  STEREO-B (STEREO-A)  was
68$\arcdeg$ East of (64$\arcdeg$ West of) the Sun--Earth line 
on 31 December 2009.

We compare EUVI with COR-1/2 data to associate flares with CMEs.
The typical temporal resolution of
COR-1 (COR-2) data is 5\,--\,10 (15\,--\,30) minutes.  
Other data used in this
study include LASCO on SOHO, WAVES \cite{Bougeret95}, and SWAVES
\cite{Bougeret08} on \textit{Wind} and STEREO, respectively.

\subsection{Online Flare List and Analysis}
 
The present study comes largely from the online EUVI flare list 
with extensive movies 
% that
we maintain at \urlurl{secchi.lmsal.com/EUVI/MOVIES\_FLARES\_CMES}.  The
list goes beyond 2010, but its main purpose is to provide movies and
plots for flares during 2007\,--\,2009.  The front page of the list at
the above address shows basic parameters of the flares, including
simplified ranks of the four CME-related attributes as described below.
More substantial material is contained
in the page for each event that is located one layer deeper
and accessible by clicking the start time in the front page.  
For example, ``2007/12/31 00:37'' leads to 
\urlurl{secchi.lmsal.com/EUVI/MOVIES\_FLARES\_CMES/event\_20071231\_00.html}
for the flare \textsf{SOL2007-12-31T01:11}\protect \footnotemark. 
Let us call it the event page.

\footnotetext{For SOL identification convention, see {\it Solar Phys.}
   {\bf 263}, 1, 2010. doi:\doiurl{10.1007/s11207-010-9553-0}}

\begin{figure}[!h]    %%%%%%%%%%%%%%%%%% FIGURE 2
 \centerline{\includegraphics[width=0.99\textwidth,clip=]{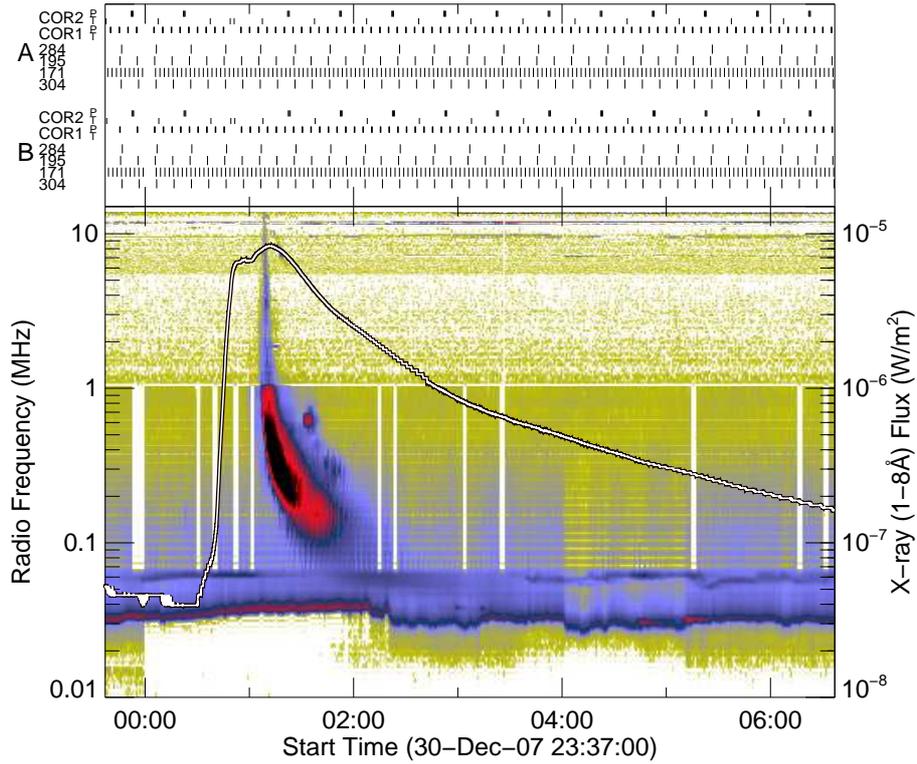}}
% {./secchi_waves_xrs200712302337.ps}}
              \caption{GOES light curve of the C8.3 flare 
\textsf{SOL2007-12-31T01:11}\protect, plotting over the
                decametric-hectrometric (DH) radio dynamic spectrum by
                \textit{Wind}/WAVES.  Data coverage of EUVI, COR-1 and COR-2
                are shown on top separately for STEREO-A and STEREO-B.
}
   \label{secchi_coverage_20071231}
\end{figure}

In the event page, the first item to look at 
is the plot that shows data coverage plotted with respect
to the GOES X-ray light curve (Figure~2).  
% This appears by clicking the start time in the top left part of the event page.  
We use this to find if
the lack of a CME or CME-related signatures is 
due to unexpected long data gaps or poor cadence.  
Figure~2 also shows a radio dynamic spectrum from
\textit{Wind}/WAVES.  This readily tells us if the flare is accompanied by 
a Decametric-Hectometric (DH) Type III burst, which is due to nonthermal
electrons that escape into the heliosphere along open field lines.  
We find that the
flares associated with a CME are often accompanied by a Type III
burst (see Section 4), 
which may bear additional information on the interface between the CME
expansion and neighboring open field regions 
(see \citeauthor{Yan06}, \citeyear{Yan06}).  We rank the flare 
in terms of the association with a Type III burst
in the following manner: i) No Type
III burst, ii) Type III burst delayed with respect to the flare
impulsive phase, iii) Type III burst that is weak especially at high frequencies
($>$1~MHz), although appearing in the impulsive phase, iv) Clear Type
III burst up to the highest frequency (13.75~MHz) of \textit{Wind}/WAVES,
appearing in the impulsive phase.  In this scheme the event shown in
Figure~2 is ranked ii),
meaning that the Type III burst is delayed with respect to the
impulsive phase.
  It is possible that the timing of the Type III burst
with respect to the impulsive phase may hold useful information on 
the role of magnetic reconnection in the overall CME processes. 
Rank iii) flares suggest that the source region may be occulted, or that 
the electrons responsible for the Type III burst are accelerated only in
the high corona, 
such as above $\approx$2\,R$_{\sun}$ in heliocentric distance.

The event page is populated with links to movies. 
% in Javascreipt (J) and Mpeg (M).  
Inspecting these movies is an essential part of the
present study, which nevertheless accommodates some quantitative analysis
as well. 
First, we make partial frame (PF) movies of 
the flare region in all of the available wavelengths, 
which consist of 640$\times$640 imagesin full resolution,
spanning an approximately 17$\arcmin \times$17$\arcmin$ field-of-view (FOV). 
The FOV is manually adjusted to capture the CME signatures as well as the
flare center.

In all wavelengths but 304~\AA, the PF movies contain intensity images
and both running and base (pre-event subtracted) difference images.  
The inclusion of base-difference images is specifically to find
long-term coronal dimming, which does not come out in more commonly
used running-difference images \cite{Attrill07}.  
In order to make base-difference
images, we try to take out solar rotation, using the SolarSoft routine 
{\tt drot\_map.pro}, 
which assumes differential rotation at the zero height from the
photosphere.  
This assumption is obviously not correct, and the inaccuracy builds up
with time, so the length of the movies should in principle not exceed
six hours for regions close to central meridian.  The off-limb portion
is not de-rotated because we do not know the longitude of each pixel. 
This can also produce artificial
dimming or brightening caused by 
the interplay between the unknown intensity distribution and solar rotation.

\begin{figure}    %%%%%%%%%%%%%%%%%% FIGURE 3
 \centerline{\includegraphics[width=0.99\textwidth,clip=]{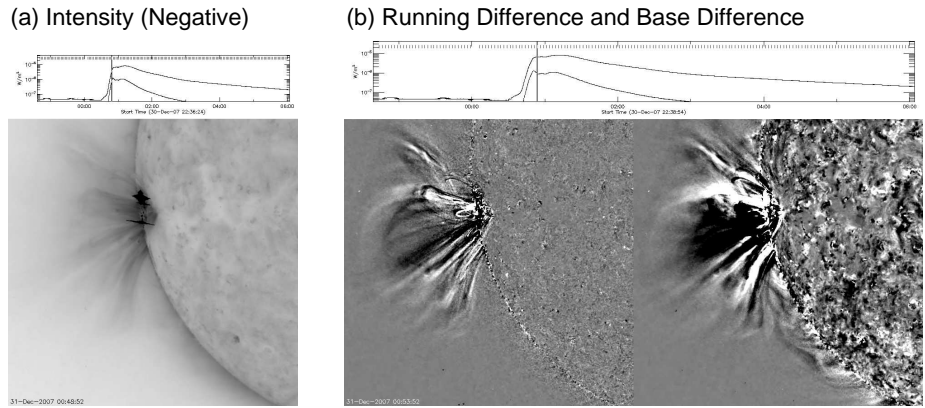}}
% {./sample_euvi_snapshots_rev.eps}}
              \caption{Snapshots from the online PF movies of the event shown in
                Figure~2.  These are of 171~\AA\ images from EUVI on
                STEREO-B
                in intensity (a) and running and base
                difference (b).  
}
   \label{snapshot_pf_171b_20071231}
\end{figure}

The movies incorporate GOES X-ray light curves on which the times of the images
are indicated.  This helps us find the temporal relation of the CME
signatures with the flare development.  
In Figure~3 we show snapshots from the PF movies of
the event plotted in Figure~2.  
Figure~3(a) shows an intensity images in 171~\AA. 
Figure~3(b) shows a running-difference image on the
left and base-difference image on the right.   
The running (base)-difference image
reflects the time difference of ten minutes (two hours).  The dimming
beneath the front is clearer in the base difference image. 

These PF movies allow us to rank the flare 
in terms of how eruptive it is in the lower corona:
i) No eruption, ii) Confined eruption, iii) Jet-like
eruption, and iv) Loop-like eruption.  We tend to connect rank iii) events
to narrow CMEs and rank iv) events to 
normal (flux-rope) CMEs. (\citeauthor{PFChen11} \shortcite{PFChen11}
  discusses the possible distinction of narrow and normal CMEs). The wavelength at
which the eruption is seen most clearly depends on the event,
probably reflecting different plasma temperature distributions 
in different eruptions.

\begin{figure}    %%%%%%%%%%%%%%%%%% FIGURE 4
%  \centerline{\includegraphics[width=0.99\textwidth,clip=]{./20090213_0500_b_195_pf_640.eps}}
 \centerline{\includegraphics[width=0.99\textwidth,clip=]{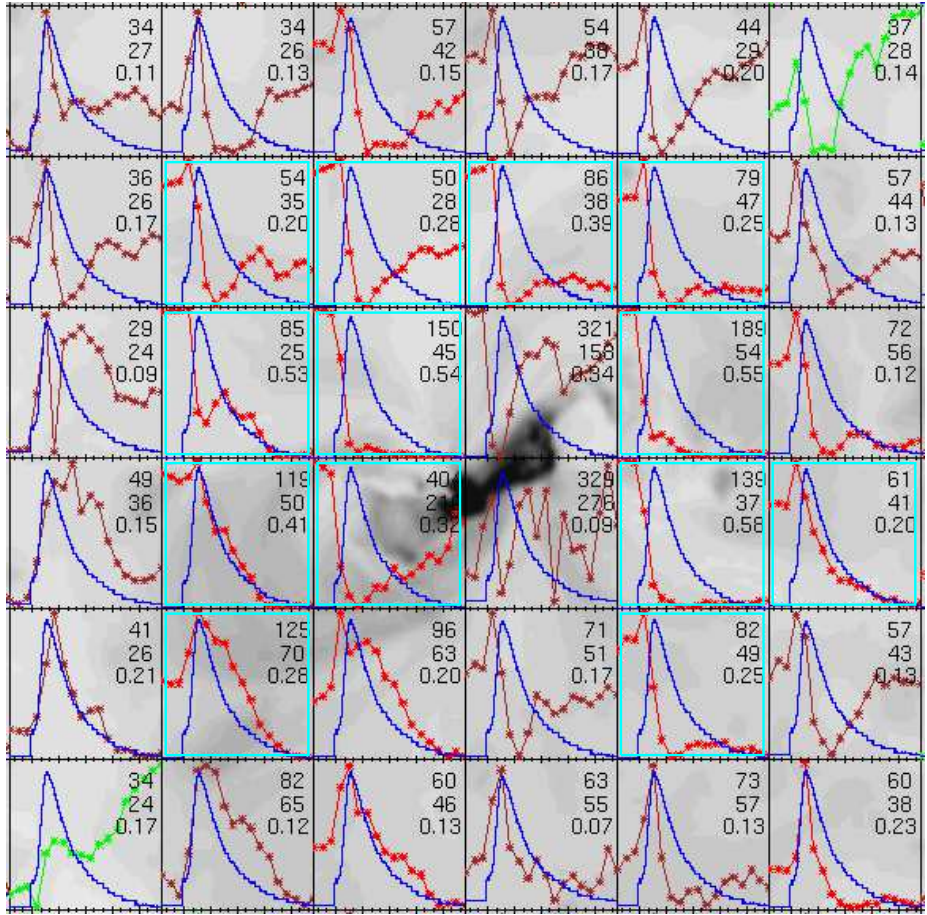}}
% {./lc_macro_pixels_only_central_20090213_0500.eps}}
              \caption{Mosaic of the temporal variation plots of the
                macropixel-averaged flux $\overline{F}$.  It is normalized to
                the minimum and maximum values of $F$ in each macropixel.
                The background image is a representative
                full-resolution image displayed on the matching scale
(one macropixel$\approx$ 64$\arcsec \times$ 64$\arcsec$).
The flux variations are color coded
on the basis of $\Delta$$F$, which is the difference between the last
and first points, $F_{\tt{last}} - F_{\tt{first}}$, that is, 
light green, red, and brown correspond, respectively to
$\Delta$$F>$0, $\Delta$$F<$0.  The
normalized GOES 1\,--\,8~\AA\ light curve is plotted in blue.  The
numbers give the maximum, minimum, and normalized difference of $F$.  
This example comes from the EUVI/STEREO-B
195~\AA\ images of the B2.3 flare of \textsf{SOL2009-02-13T05:47}.  It is the
central part of  
\protect
\urlurl{secchi.lmsal.com/EUVI/MOVIES\_FLARES\_CMES/lc\_macro\_pixels/20090213\_0500\_b\_195\_pf\_640.gif}.
The 13 macropixels with boxes in cyan show substantial dimming.
}
   \label{lc_macro_pixels_only_central_20090213_0500.eps }
\end{figure}

In order to characterize coronal dimming beyond visual inspection of
the movies, we average the 640$\times$640 full-resolution PF images into 
16$\times$16 macropixels, 
and follow the temporal variations of the flux in each macropixel.  
We plot the flux variations on top of 
a representative full-resolution image.
These plots in all the available wavelength bands are linked in
the event page.  We color-code the flux curves to easily find  
macropixels that show substantial dimming.  They
are plotted in light green (red) if the flux increases
(decreases) in the plotted time interval.  
Those plotted in brown indicate that the
difference between the first and last points is small. 
The flux variations can be
compared with the normalized GOES light curve (in blue).
The maximum flux $F_{\tt{max}}$, minimum flux $F_{\tt{min}}$ and 
normalized difference ($F_{\tt{max}} - F_{\tt{min}}) / ( F_{\tt{max}} - F_{\tt{min}})$
are shown.
As an example, we show in Figure~4 the central part of 
one of the plots.  
% An example is given in Figure~4.  
% We throw away the right-most or left-most column 
% in order to avoid the effect of de-rotating the images.  
Here we look for macropixels that undergo dimming by
more than 10\,\% of the pre-flare level and stay dimmed for more than an
hour.  In Figure~4, these macropixels are marked by boxes in cyan.
We count macropixels with $>$10\,\% dimming 
in different wavelengths depending on
their availability and use the numbers to rank the flare 
in terms of the magnitude of dimming:
i) 0\,--\,5 macropixels, ii) 6\,--\,10 macropixels, iii) $>$10
macropixels in one wavelength band, iv)  $>$10
macropixels in two wavelength bands and v)  $>$10
macropixels in three wavelength bands.  The event used in Figure~4 is
ranked to be v), since it has more than ten macropixels with $>$ten\,\%
dimming in the 171~\AA, 195~\AA, and 284~\AA\ bands.
Note that after August 2009,
it has become difficult to assign ranks iv) and v) because of the lack of
images at a reasonable cadence at 171~\AA\ or 284~\AA.  Despite the
concern that dimming at only one wavelength may represent temperature
effects rather than mass evacuation due to a CME, 
we have leanred that
deep dimming at 195~\AA\ is usually a reliable coronal signature of CMEs.
Therefore we consider rank iii) to represent 
CME-related dimming as significant 
as ranks iv) and v).

\begin{figure}    %%%%%%%%%%%%%%%%%% FIGURE 5
 \centerline{\includegraphics[width=0.99\textwidth,clip=]{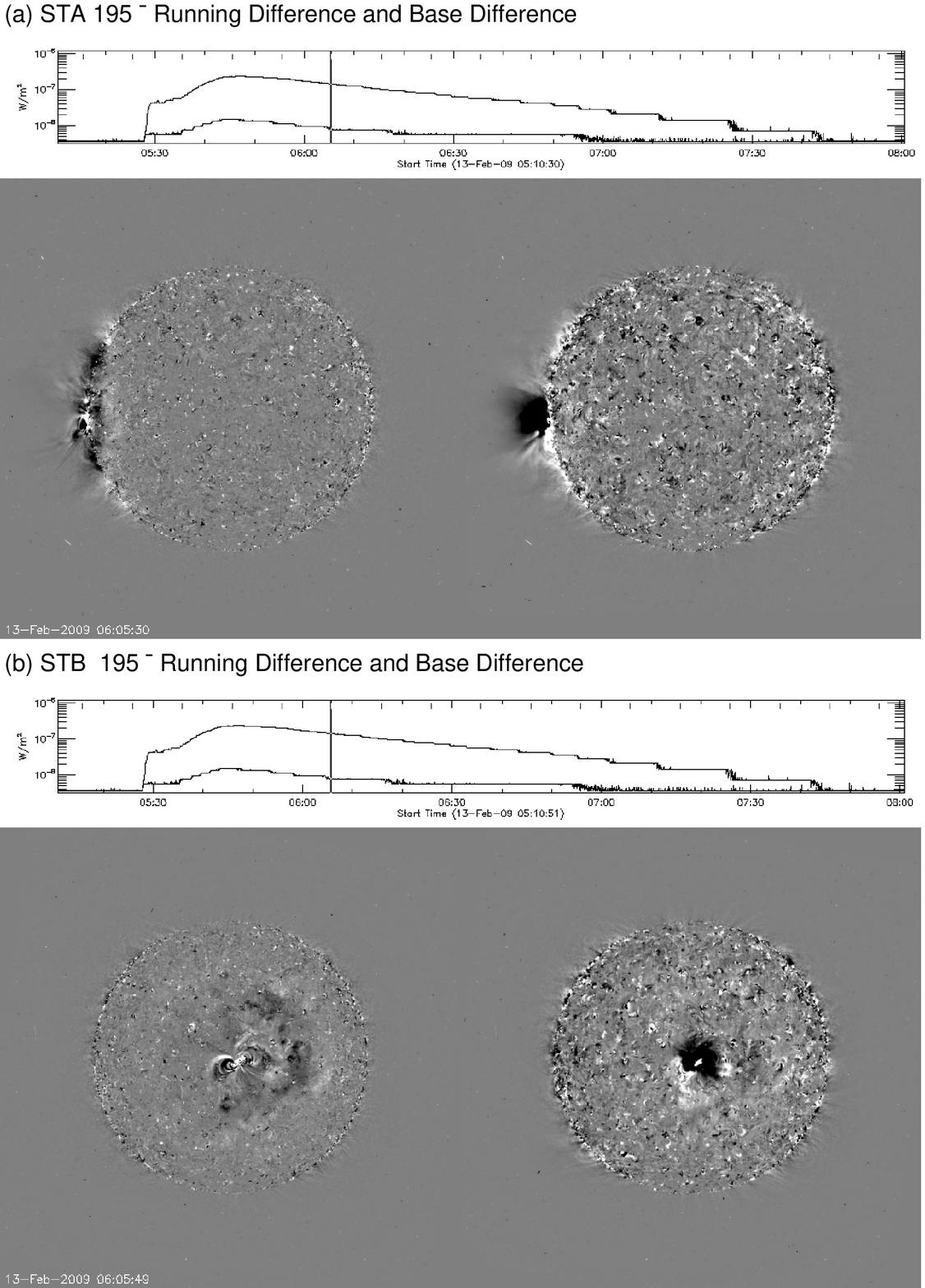}}
% {./sample_euvi_snapshots_fd_20090213_05_rev.eps}}
              \caption{Snapshots from the movies of full-disk
                difference images in 195~\AA.  In each panel, the image
                on the left is the difference of images ten
                minutes apart, whereas the image on the right
                is the pre-event subtracted image.  In this case
                the time difference is 45 minutes.
}
   \label{sample_euvi_snapshot_fd_20090213_05}
\end{figure}

Next, we look for EIT or EUV waves and study their relations with CMEs. 
See \citeauthor{Patsourakos12} \shortcite{Patsourakos12} for the most
recent review on EUV waves.  
To determine whether the flare is associated with
EUV waves, the event page
also carries movies of full-disk (FD) images in the 195~\AA\ band.
EUV waves usually appear most pronounced in
195~\AA\ images, and they often propagate beyond the FOV of the
PF images extracted around the flare region.
Snapshots from FD movies are shown in Figure~5.  This particular 
event (also used for Figure~4) was observed by STEREO-A and STEREO-B 
almost 90 degrees apart, 
% in quadrature, 
which provide both the disk and limb views of 
the eruption.  Thus this event has been extensively studied 
({\it e.g.} \citeauthor{Patsourakos09}, \citeyear{Patsourakos09}; 
\citeauthor{Kienreich09}, \citeyear{Kienreich09}; 
\citeauthor{Cohen09}, \citeyear{Cohen09}).
We rank the flare in terms of the association with an
EUV wave on the basis of visual
inspection of these movies.  The ranking is: i) No wave, ii) Marginal
wave, iii) Smaller-scale wave (propagating to a distance 
less than 0.5\,R$_{\sun}$,
iv) Large-scale wave.  The example in Figure~5 is clearly rank iv).
Note that this ranking is simply based on the appearance of
  the propagating front.  It could be more physics-based if we used
  the measured speed of the front.  
\citeauthor{Warmuth11} \shortcite{Warmuth11} proposed a scheme to
differentiate the origin of EUV waves on the basis of the kinematics.
However, we could not identify similar distinct subgroups in our sample.

\begin{figure}    %%%%%%%%%%%%%%%%%% FIGURE 6
 \centerline{\includegraphics[width=0.99\textwidth,clip=]{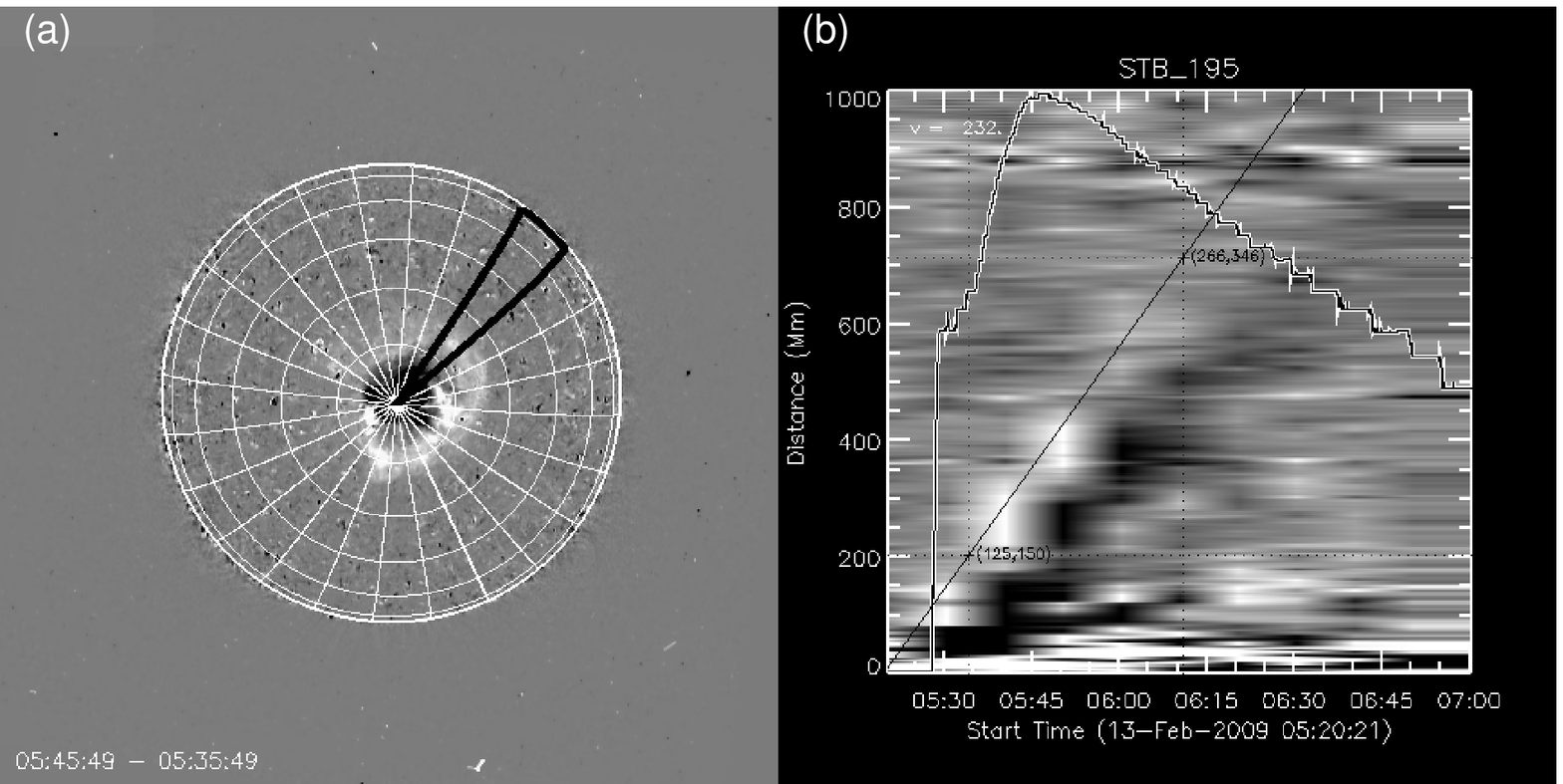}}
% {./click_space_time_stb195_vel_sector16_20090213_rev.eps}}
              \caption{The intensity profile is measured in the 24 sectors along
                great circles with the origin at the center of the
                eruption (a). The
                running-difference distance--time plot (b)
is for the sector
                in black in panel (a).
}
   \label{click_space_time_stb195_vel_sector16_20090213.eps}
\end{figure}

\begin{table}

\caption{Coronal waves observed by EUVI during March 2007\,--\,December 2009 
}
\label{W-simple}

\tabcolsep 5.8pt
\begin{tabular}{rrrrrrrr}     % define the column alignment
                           % l: left, c: center, r: right
  \hline                   % horizontal line
 
Date and Time & GX & Loc. & A or B$^{1}$ & AR$^{2}$ & $v_{195}$$^{3}$ & $v_{171}$$^{3}$ & CME$^{4}$ \\ 
\hline
 8 May 2007 05:51 & B1.2  & S05 E90 & B &   N-AR & 419  & 366 & v \\
16 May 2007 17:19 & C2.9  & N00 E35 & B &  10956 & 283  & 274 & iv \\
19 May 2007 12:48 & B9.5  & N00 E03 & A &  10956 & 341  & 521 & v \\
20 May 2007 04:52 & B6.7  & N00 W07 & A &  10956 & -195 & 275 & v \\
22 May 2007 14:30 & B3.9  & N03 W38 & A &  10956 & 322  & -217 & v \\
23 May 2007 07:15 & B5.3  & N03 W53 & B &  10956 & 357  & -307 & v \\
 3 Jun 2007 09:23 & C5.3  & S08 E61 & B &  10960 & 443$^{5}$ & 271$^{5}$ & iv \\
 6 Aug 2007 01:45 & B4.2  & S07 E44 & B &  10966 & 276  & NA & iii \\
 6 Aug 2007 09:05 & C1.5  & S05 E41 & A &  10966 & 392  & 365 & iv \\
 6 Aug 2007 15:24 & C1.1  & S05 E38 & B &  10966 & 295  & 167 & iv \\
 7 Dec 2007 04:35 & B1.4  & S05 W06 & A &  10977 & 289  & 277 & v \\
 8 Dec 2007 17:08 & N-FL  & S05 W27 & A &  10977 & 269  & 192 & v \\
31 Dec 2007 00:37 & C8.3  & S10 E106 & B &  10980 & 241  & 248 & v \\
 2 Jan 2008 06:51 & C1.1  & S06 E72 & B &  10980 & 88  & 108 & v \\
 7 Jan 2008 02:26 & B1.2  & N38 W02 & A &  10981 & 366  & 315 & ii \\
25 Mar 2008 18:36 & M1.7  & S11 E80 & B &  10989 & 435  & -447 & v \\
 5 Apr 2008 15:36 & A5.8  & S08 W105 & A &  10987 & 408$^{5}$  & -411$^{5}$ & v \\
16 Apr 2008 19:37 & A5.2  & N30 W04 & A &  10990 & 335  & 257 & ii \\
26 Apr 2008 13:41 & B3.8  & N10 E10 & B &  N-AR & 275  & -208 & v \\
 4 Nov 2008 03:17 & C1.0  & N37 W47 & A &  11007 & 260  & 201 & v \\
11 Dec 2008 05:22 & B2.6  & S26 W61 & A &  11009 & 305  & 370 & iii \\
11 Dec 2008 09:22 & C1.4  & S25 W88 & A &  N-AR & 214  & 287 & ii \\
10 Feb 2009 23:00 & B1.3  & S06 E75 & B &  11012 & 262  & -210 & v \\
12 Feb 2009 16:10 & B4.1  & S06 E58 & B &  11012 & 250  & 402 & iii \\
13 Feb 2009 05:35 & B2.3  & S06 E46 & B &  11012 & 232  & 267 & v \\
27 Feb 2009 07:10 & A3.0  & N27 W10 & A &  11012 & 182  & 200 & iii \\
 1 Oct 2009 02:47 & B7.0  & N23 W80 & A &  11027 & 221  &  - & ii \\
15 Dec 2009 23:10 & N-FL  & S33 W53 & A &  N-AR & 195  &  - & i \\
16 Dec 2009 01:02 & C5.3  & N30 W06 & A &  11035 & 434  &  - & v \\
21 Dec 2009 23:34 & B4.1  & S27 W44 & A &  11036 & 348  &  - & iii \\
22 Dec 2009 04:50 & C7.2  & S27 W46 & A &  11036 & 403  &  - & v \\
22 Dec 2009 15:10 & C1.1  & S27 W51 & A &  11036 & 335  &  - & i \\
22 Dec 2009 20:23 & C1.3  & N16 W51 & A &  11038 & 302  &  - & i \\
23 Dec 2009 10:09 & C6.4  & S28 W62 & A &  11036 & 321  &  - & iii \\

\hline
\end{tabular}

1. A or B indicates which EUVI data are used to measure the speed.  A
(B) means EUVI on STEREO-A (B). \\
2. AR number assigned by NOAA.  N-AR means no active region or outside
numbered active regions. \\
3. The minus sign indicates that the front appears in negative.  \\
4. The CME rank i)\,--\,iv).  See the definition given in Section 3.3.\\
5. Measured along the limb.  \\

\end{table}

Irrespective of flares, we have surveyed all of the EUVI data between
March 2007 and December 2009 and found 34 EUV waves, including small
ones that correspond to rank iii) as mentioned above. They are given
in Table~2.  Most of them are associated with flares, but a few minor
ones are not associated with any increase in the GOES X-ray flux.  For
each wave, we have measured the distance of the front 
from the center of the eruption or flare along the great
circle, and studied the speed of the front
corrected for the curvature of the solar surface.  
This technique is widely used for detecting and characterizing EUV
waves ({\it e.g.} \citeauthor{Podladchikova05}, \citeyear{Podladchikova05}).
We make distance--time plots for 24 equally-spaced sectors. 
The fronts appear as bright ridges in running-difference distance--time plots
(Figure~6).
In many cases, the ridges appear
only in a small number of sectors.  We choose the sector that shows the
front clearly and yields the highest speed.  This speed is entered in
the sixth (195~\AA) and seventh (171~\AA) columns of Table~2.  
The same EUV waves were often observed by EUVI on both
STEREO-A and STEREO-B while their separation was small.  
In these cases we give the measured speed
from one spacecraft only, after confirming its consistency with that from
the other spacecraft within the uncertainties.  The uncertainties 
are estimated to be as large
as $\approx$100~km~s$^{-1}$ on the basis of multiple attempts to trace the
ridges.

The typical
195~\AA\ cadence was ten minutes until August 2009, which is comparable
to that of EIT, so the measurement of the speed of the front in EUVI
data is nowhere as refined as in the 12~second data from 
the {\it Atmospheric Imaging Assembly} 
(AIA: \citeauthor{Lemen12}, \citeyear{Lemen12}) onboard the 
{\it Solar Dynamics Observatory} (SDO: \citeauthor{Pesnell12},
\citeyear{Pesnell12}).  
As of May 2013, more than 30 articles have been
published on AIA observations of EUV waves 
(mostly case studies) as cited by 
\citeauthor{Nitta13b} \shortcite{Nitta13b}, who conducted an ensemble
study of large-scale coronal propagating fronts observed by AIA.
% Examples of studying the kinematics of EUV
% waves using AIA data can be found in \citeauthor{Liu10} \shortcite{Liu10} and
% \citeauthor{Nitta13b} \shortcite{Nitta13b} among several others.
A few limb events in Table~2 do not show EUV waves on disk, but we
clearly see indications of large-scale disturbances over the limb.  We
measure their speed along the limb at the height of
0.15\,R$_{\sun}$, 
which is consistent with the
stereoscopic estimates of the height of EUV waves 
(see \citeauthor{Patsourakos09b}, \citeyear{Patsourakos09b};
\citeauthor{Kienreich09}, \citeyear{Kienreich09}), 
similar to the work by \citeauthor{Downs11} \shortcite{Downs11} 
and \citeauthor{Liu12} \shortcite{Liu12}.  Despite
large uncertainties, we confirm that the speed of the EUV waves during
solar minimum basically conforms to the range of those observed by EIT
\cite{Thompson09}.  We also find examples of faster fronts in
171~\AA\ than in 195~\AA\ due to a better cadence as argued by
\citeauthor{Long08} \shortcite{Long08}.  However, we point out that in
many cases the EUV wave is not as well defined at 171~\AA\ as at 195~\AA.  In
some cases it appears as a dark ridge as marked in Table~2, probably
as a result of heating,
as discussed by 
\citeauthor{Wills_Davey99} \shortcite{Wills_Davey99},
who analyzed data from both EIT and 
the {\it Transition Region and Coronal Explorer} 
(TRACE: \citeauthor{Handy99}, \citeyear{Handy99}).

\subsection{Associating Flares with CMEs}

Lastly, we determine whether the flare is associated with a CME, which 
involves ranking the outflow or the lack thereof in coronagraph data. 
We first check the CDAW LASCO CME catalog (\urlurl{cdaw.gsfc.nasa.gov/CME\_list/},
see \citeauthor{Yashiro04}, \citeyear{Yashiro04}).  Using the spatial
and temporal proximities, it is possible to determine whether the flare is
associated with a CME.  However, there are many ambiguous events
because LASCO's FOV starts only at $>$2.2\,R$_{\sun}$ from disk center
and the cadence is typically between 10 and 60 minutes.  The
ambiguities in many of these cases may be resolved by 
STEREO/COR-1 data because their
cadence is
typically five (or ten) minutes and they cover the heliocentric
distance of 1.4\,--\,4.0\,R$_{\sun}$.  If a CME is associated with the
flare, we expect it to emerge in COR-1 data before it is
observed by COR-2 or LASCO. 

\begin{figure}    %%%%%%%%%%%%%%%%%% FIGURE 7
 \centerline{\includegraphics[width=0.99\textwidth,clip=]{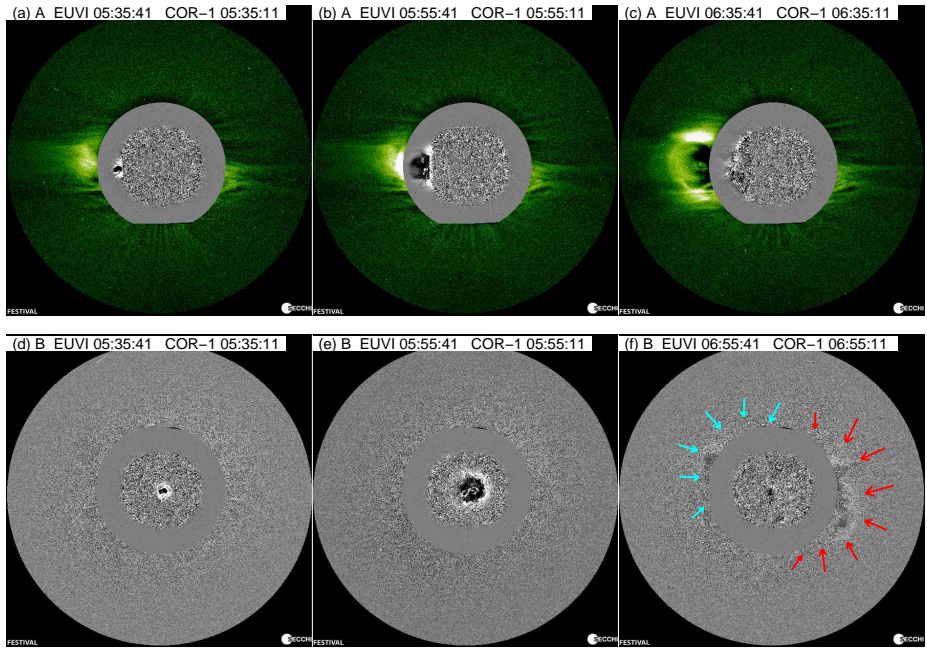}}
% {./festival_20090213_rev.eps}}
              \caption{The eruption that
                appears in EUVI images as a wave or propagating front
                is observed as a CME in COR-1 images.
                Panels (a)\,--\,(c) come from STEREO-A, providing a
                limb view.  Panels (d)\,--\,(f) come from STEREO-B,
                providing a disk view. The EUVI images
              are given as running-difference images.  The COR-1
              images from STEREO-A (STEREO-B) are
              intensity (running-difference) images.  In panel (f)
              the CME front is indicated by red and cyan arrows. }
   \label{festival_20090213}
\end{figure}

We extensively use the convenient software 
FESTIVAL \cite{Auchere08} as included in SolarSoft \cite{Freeland98} 
to make EUVI and COR-1 (and COR-2) composite movies.  
Snapshots from such movies are shown in
Figure~7, where we first note an EUV wave and then a CME.
The CME is more easily
seen in the limb view from STEREO-A.  From STEREO-B, this event occurred 
close to disk
center, making it hard to observe the CME.  However, difference movies
clearly show the CME coming from the west and northeast limbs.
It is noted that some outflows in the COR-1 FOV disappear before reaching
the COR-2 FOV.  Indeed, 
these outflows are usually not found by
the Computer Aided CME Tracking (CACTus: \citeauthor{Robbrecht04},
\citeyear{Robbrecht04}; \citeauthor{Robbrecht09a}
\citeyear{Robbrecht09a}) as applied to COR-2 data 
(\urlurl{secchi.nrl.navy.mil/cactus}). 

Fully characterizing or modeling individual CMEs is beyond the
scope of the present work.  Instead, we rank the outflows in
coronagraph data in the following manner:  i) No outflow, ii) Outflow that
disappears within 3\,R$_{\sun}$, iii) Narrow CME, typically narrower than
30$\arcdeg$,  iv) CME detectable up
to 10\,R$_{\sun}$, 
v) CME detectable beyond
10\,R$_{\sun}$.  Defining CMEs as eruptions of flux ropes into the
heliosphere, this ranking is meant to distinguish CMEs from 
other, but possibly related, physical processes
such as waves (rank ii)) and jets (rank iii)).  
In our classification, only rank v) may represent
real CMEs, but 
they may be grouped into rank iv) by the interplay
  between the direction of motion of the CME (with respect to the
  plane of the sky) and the limited sensitivity of the instrument.
% the limited sensitivity may put some CMEs in rank 4.
Therefore we consider rank iv) and v) events to be real CMEs.
The CME ranking comes primarily from the CDAW CME Catalog and CACTus,
but we conduct our own measurements when needed.
An example of the CME ranking is shown in the last column of 
Table 2 for EUV waves.

\section{CME-Associated Flares}

\begin{table}

\caption{Flares associated CMEs during March 2007\,--\,December 2009 
}
\label{T-simple}

\tabcolsep 3.8pt
\begin{tabular}{crrccrllcccc}     % define the column alignment
                           % l: left, c: center, r: right
  \hline                   % horizontal line
% \multicolumn{2}{c}{Flare} & \multicolumn{2}{c}{AR}&
% \multicolumn{3}{c}{CME} &  &  & \\ 
 
1 & 2 & 3 & 4 & 5 & 6 
& 7 & 8 & 9 & 10  & 11 \\
Date and Time & A09 & GX & Loc. & AR & $v$ & 
$\theta$ & D & E & W & III  \\
\hline
25 Mar 2007 13:22 &     & A4.2  & S13 W77 & 10947 & 199 & 39 & iv & iv & i & i \\
 2 May 2007 18:05 &     & B3.8  & S15 W14 & 10953 & 275 & 95 & v & iv & i & i \\
 8 May 2007 05:51 &     & B1.2  & S05 E90 & N-AR & 462 & 52 & iv & iv & iv & iii \\
15 May 2007 09:28 &     & B1.0  & N04 E57 & 10956 & 163 & 53 & ii & iii & i & iii \\
15 May 2007 15:27 &  40 & C1.0  & N00 E50 & 10956 & 190 & 65 & i & iii & i & iv \\
15 May 2007 18:02 &  41 & B3.2  & N00 E48 & 10956 & 491 & 134 & v & iv & ii & iv \\
16 May 2007 17:19 &  43 & C2.9  & N00 E35 & 10956 & 371 & 75 & ii & iv & iv & iv \\
19 May 2007 12:48 &  46 & B9.5  & N00 E03 & 10956 & 958 & 106 & v & iv & iv & iv$^{*}$ \\
20 May 2007 04:52 &     & B6.7  & N00 W07 & 10956 & 275 & 92 & v & iv & iv & iii \\
22 May 2007 00:25 &     & B1.2  & N03 W20 & 10956 & 260 & 81 & ii & iv & ii & iii \\
22 May 2007 14:30 &     & B3.9  & N03 W38 & 10956 & 544 & 108 & v & iv & iv & iv$^{*}$ \\
23 May 2007 07:15 &  48 & B5.3  & N03 W53 & 10956 & 679 & 90 & iii & iv & iv & iv$^{*}$ \\
26 May 2007 16:34 &     & B1.8  & N03 W105 & 10956 & 197 & 79 & iv & iv & i & i \\
 1 Jun 2007 22:17 &  59$^{+}$   & C2.8  & S08 E82 & 10960 & 584 & 75 & ii & iv & i & i \\
 3 Jun 2007 09:23 &  72 & C5.3  & S08 E61 & 10960 & 467 & 71 & v & iv & iv & iv$^{*}$ \\
16 Jul 2007 00:50 &     & B2.9  & N03 W57 & 10964 & 307 & 34 & iv & iv & i & i \\
 6 Aug 2007 09:05 & 132 & C1.5  & S05 E41 & 10966 & 379 & 60 & i & iv & iv & iv$^{*}$ \\
 6 Aug 2007 15:24 & 133 & C1.1  & S05 E38 & 10966 & 207 & 54 & iii & iv & iv & iv \\
18 Aug 2007 11:24 &     & B3.3  & S08 E106 & 10969 & 516 & 46 & ii & iv & i & iv \\
24 Aug 2007 02:49 &     & A4.6  & S09 E18 & N-AR  & 183 & 118 & iii & iv & i & iv \\
 7 Dec 2007 04:35 &     & B1.4  & S05 W06 & 10977 & 284 & 55 & iv & iv & iv & iv \\
31 Dec 2007 00:37 & 175 & C8.3  & S10 E90 & 10980 & 995 & 164 & v & iv & iv & ii$^{*}$ \\
 2 Jan 2008 06:51 & 177 & C1.1  & S06 E72 & 10980 & 676 & 143 & v & iv & iv & i \\
25 Mar 2008 18:36 & 183 & M1.7  & S11 E80 & 10989 & 1103 & 112 & v & iv & iv & iv$^{*}$ \\
 5 Apr 2008 15:39 &     & A5.8  & S08 W105 & 10987 & 962 & 209 & iv & iv & iv & i \\
26 Apr 2008 13:41 &     & B3.8  & N10 E10 & N-AR  & 515 & 281 & v & iv & iv & iv$^{*}$ \\
17 May 2008 10:05 &     & B1.7  & S08 E26 & N-AR     & 630 & 123 & ii & iv & i & iv \\
14 Jun 2008 23:32 &     & A4.7  & S10 W01 & 10998  & 243 & 73 & iv & iv & i & iii \\
 4 Nov 2008 03:17 & 185 & C1.0  & N37 W47 & 11007  & 732 & 66 & ii & iv & iii & iv \\
10 Feb 2009 23:00 &     & B1.3  & S06 E75 & 11012  & 312$^{\dagger}$ & 56$^{\dagger}$ & iv & iv & iv & iv \\
13 Feb 2009 05:35 &     & B2.3  & S06 E46 & 11012  & 328$^{\dagger}$ & 44$^{\dagger}$ & v & iv & iv & ii \\
31 Oct 2009 15:12 &     & B3.0  & N17 W94 & 11029  & 499 & 58 & iii & i & i & i \\
13 Dec 2009 09:17 &     & B1.4  & N20 E26 & 11034  & 406 & 75 & iii & iv & ii & ii \\
16 Dec 2009 01:02 &     & B1.4  & N30 W02 & 11035  & 276 & 360 & v & iv & iv & i \\
22 Dec 2009 04:50 &     & C7.2  & S27 W46 & 11036  & 318 & 47 & iii & iv & iv & iv \\

\hline
\end{tabular}

\noindent
1: Flare onset time.  2: Event number in Aschwanden {\it et al.}
(2009).  3: GOES flare class.  4: Location in heliographic
coordinates.
5: NOAA active region number.  N-AR stands for regions without NOAA
active regions.  6: CME linear speed (km~s$^{-1}$) from the CDAW CME catalog unless otherwise
noted.
7: CME width (degrees) from the CDAW CME catalog unless otherwise
noted.  8: Rank for dimming.
9: Rank for eruptions. 10: Rank for EUV waves.  11: Rank for the
association with Type III bursts.

\noindent
$^{+}$ Aschwanden {\it et al.} (2009) lists the interval
21:35\,--\,22:35~UT represented by the preceding M2.1 flare, which is
not associated with a CME. 

\noindent
$^{\dagger}$ CME parameters taken from CACTus. 

\noindent
$^{*}$ Metric Type II burst.

\end{table}

In Table~3, we give the results of the analysis described in Section
3.2.  Rather than listing all of the flares we studied, namely 11
M-class, 95 C-class and tens of B-class and less intense flares, we
show only those flares associated with a CME with the CME rank of iv) and v).  
The mean speed and width of the CME are given in the sixth and seventh
columns.  
Most of them are taken from the CDAW CME Catalog, and
  they are not corrected for the angle between the plane of the sky
  and the direction of the CME propagation expected for the location
  of the source region. The CME speed as observed in LASCO-C2 and -C3 
data should be taken as a lower limit, since the CME is often seen to
decelerate already in this FOV.
It may be surprising that only 35 flares are included, and more so
that a majority of them are B-class and less intense flares.  
M- and C-class flares are not as strongly associated with CMEs
found for Solar Cycle 23 (\citeauthor{Yashiro05}, \citeyear{Yashiro05}).
The association rate is much lower than the frequency of reports of
CME occurrence that were shown by
\citeauthor{Aschwanden09} \shortcite{Aschwanden09}.
In fact, Table~3 includes only 13 of the 159 overlapping flares in 
\citeauthor{Aschwanden09} \shortcite{Aschwanden09},
as indicated in the second column.   
More recently, \citeauthor{Bein11} \shortcite{Bein11}
studied statistically
the properties of 95 impulsively accelerated CMEs, mostly
flare-associated, during the overlapping
period (January 2007\,--\,May 2010).  It appears that many of them
were from 2010 (after the period that we deal with here), 
when the activity level resumed to be comparable to
that of early 2007.

Despite the low association rate of flares with CMEs, we note that all
of the four fast ($v>$ 900~km~s$^{-1}$) {\it and} wide ($\theta
>$100$\arcdeg$) CMEs during the period of interest 
are associated with a flare that is at least
identified in GOES soft X-ray light curves.  In the same period, 
there are a number of wide front-side CMEs 
not associated with a flare, but they
are never faster than 500~km~s$^{-1}$.

\begin{figure}    %%%%%%%%%%%%%%%%%% FIGURE 8
 \centerline{\includegraphics[width=0.99\textwidth,clip=]{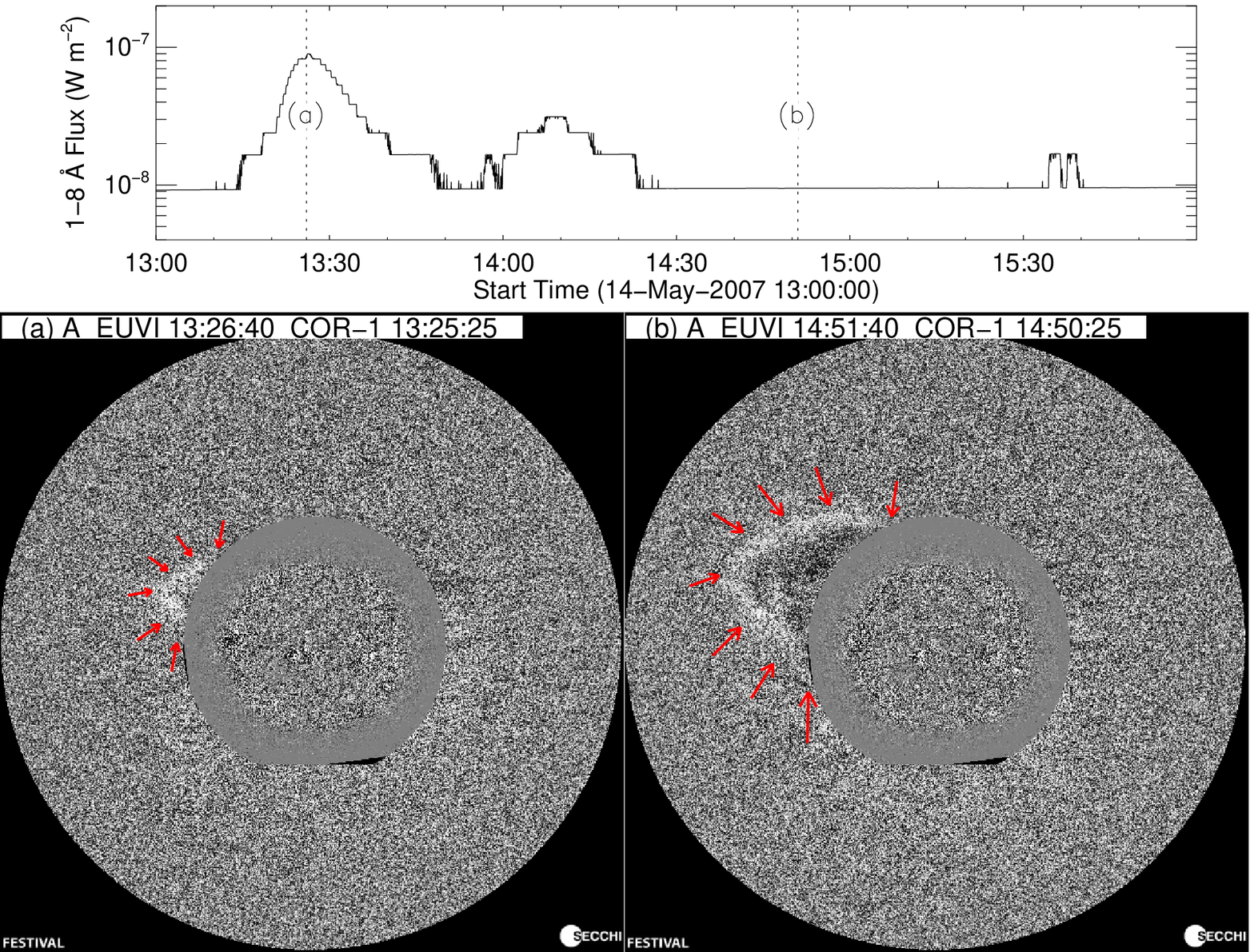}}
% {./cor1_euvi_a_20070514_13_a8flare.eps}}
              \caption{Composite COR-1 and EUVI images of the CME
                observed around the time of an A8 flare on 14 May 2007.  The CME
                front is indicated by red arrows.}
   \label{cor1_euvi_a_20070514_13_a8flare.eps}
\end{figure}

Table~3 would have been a little longer if we had included 
a few flares whose association with a CME is dubious.  
Figure~8 shows an example.   
There was an A8 flare on 14 May 2007, whose emission in EUV was
so weak that it was difficult to locate it in EUVI intensity images.  Shortly
after the flare, at 14:06~UT, LASCO observed a CME at the heliocentric
distance of 2.4\,R$_{\sun}$.  Although its linear speed was only
225~km~s$^{-1}$, 
the CME was as wide as 72$\arcdeg$ and the front was traceable to
17\,R$_{\sun}$.  Therefore, it is rank 5 in our scheme.  However, using data
from COR-1 on STEREO-A, we do not believe that the A8 flare was
associated with this CME, although we cannot rule out the
possibility.  The main reason is that the front-like structure was
already seen at $\approx$2\,R$_{\sun}$ in COR-1 data near the flare
peak (Figure~8(a)), suggesting the CME launch
from a high altitude, generally not accessible to a minor flare.  
% The EUVI
% difference image in Figure~8(a) shows a small dimming, but its offset
% from the CME front.  
The flare
region, AR~10956, was already 10$\arcdeg$ inside the east limb, but
the two STEREO spacecraft were separated by only 8$\arcdeg$ at the
time, and it was not clear if there was a large-scale structure behind
the east limb that could have accounted for this CME.

Let us examine the flare-CME association 
in terms of the signatures in non-coronagraphic
observations.  See columns 8\,--\,11 of Table~3, which show ranks of
the flares in terms of the respective observables as we introduced in
Section 3.2.  As expected, most of the CME-associated flares
appear to be highly eruptive 
in the low corona.  
The 9th column shows that 32 of 35 (91\,\%) flares are rank
iv), indicating a broad, loop-like eruption which looks unconstrained.
Note that this criterion comes from visual inspection of images and
that it may be quite subjective.
It is nevertheless straightforward to distinguish rank iv) events from
confined (failed) eruptions (rank ii)) and jets (rank iii)).
Coronal dimming (in the 8th column) with ranking of iii) or higher is seen in  
26 (74\,\%) CME-associated flares.  It is possible that our
constraints may be too strong in terms of the depth of dimming
($>$10\,\% less than the pre-flare intensity) and the number of
macro-pixels.  Note also that using the same number of pixels for the
ranking of both disk and limb flares could be misleading because we do not
know which part of the 3D coronal structure becomes dimmer due to mass
evacuation in the CME processes.  
An EUV wave is seen in a smaller number (20) of 
the CME-associated flares. 
This may be because not all CMEs go through the phase
  of fast lateral expansion, which is proposed to be an essential
  ingredient of EUV waves \cite{Patsourakos12}.
%It is generally known that an EUV wave may not be driven if the CME is launched
%slowly, such as in a filament eruption from a quiescent region.  Some
%minor flares may share properties of filament eruptions.
It is also interesting to compare Table~3 with Table~2, where 21 of 34
(62\,\%) EUV waves are shown to be associated with a clear CME.
% One EUV wave event without a CME is also 
% devoid of a noticeable flare emission.  
Lastly, although 74\,\% of the CME-associated flares show a Type III
burst, the number is much less if we limit to a clear Type III burst
that is observed at the highest frequency of \textit{Wind}/WAVES (13.75~MHz) 
and in the impulsive phase.
Flares not associated with a Type III burst
tend to be more gradual, including those associated with
a filament eruption.  For example, the B3.0 flare \textsf{SOL2009-10-31T15:24}
is associated with a slow filament eruption.  No clear Type III burst
is observed, nor is an EUV wave detected.  Therefore, it appears that
a Type III burst may be characteristic of only a certain subset of CMEs.

\begin{figure}    %%%%%%%%%%%%%%%%%% FIGURE 9
 \centerline{\includegraphics[width=0.99\textwidth,clip=]{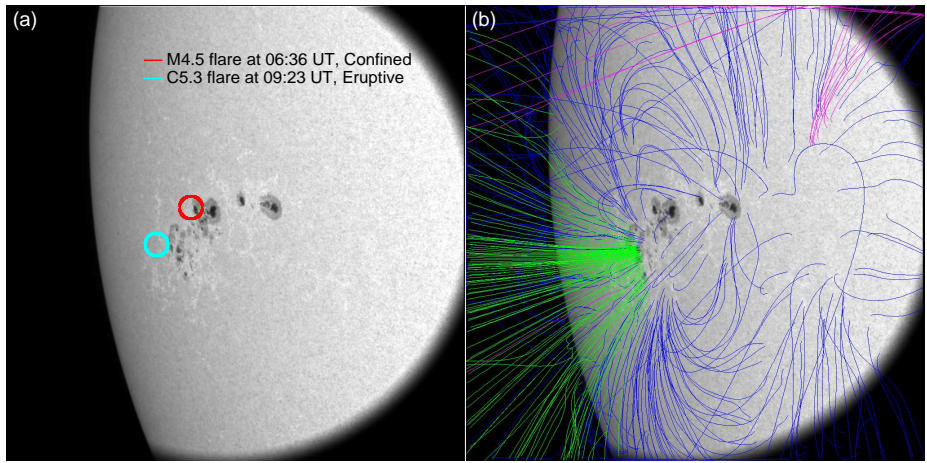}}
% {./2007060306_09_pfss_rev.eps}}
              \caption{(a) The locations of the two flares
                  on 3 June 2007, derotated, are indicated on a TRACE
                  white-light image.  (b) Field lines computed with
                  the potential field source surface (PFSS) model are
                  shown.  Blue lines represent closed field lines,
                  whereas green and pink lines open field lines
                  with positive and negative polarities, respectively,
                  at their footpoints. }
 
% The confined and eruptive flares occur at
% different locations in AR~10960.}
   \label{20070603_06_09_pfss}
\end{figure}

Finally we discuss active regions 
in terms of productivity of CME-associated flares.
% that produce CME-associated flares.  
First, we point out that several
% some 
CME-associated flares occurred in tiny
active regions or in regions without sunspots (therefore without
NOAA region numbers).
According to Table~3, as many as ten CME-associated flares occurred
in AR~10956.  They are predominantly B-class flares and there was no M-class
flare from this region (Table~1).  In contrast, 
flares in AR~10960, including 10 M-class and 17 C-class flares, 
are hardly CME-associated.   
It is speculated that the co-existence of bipoles following
and not following Hale's law \cite{Bone09} in AR~10956 may have made
the difference in terms of productivity of CME-associated flares.  This special
configuration may be more important than the sunspot area, 
the $R$ parameter  
(\citeauthor{Schrijver07}, \citeyear{Schrijver07}), and the Mount Wilson
sunspot classification scheme as given in the sixth, seventh and eighth columns of
Table~1.  Another factor that may affect whether the flare is
associated with a CME may be its location with respect to the active
region and global magnetic field.  
Figure~9 compares the locations of flares in AR~10960.
Most flares in AR~10960 are confined, 
and they occur close to the center of the region as 
encircled in red
%shown 
in Figure~9(a).  
An exceptionally eruptive flare in AR~10960 is observed on the
periphery (encircled in cyan).
% (Figure~9(b)).  
This is consistent with earlier studies
(e.g. \citeauthor{Akiyama07}, \citeyear{Akiyama07};
 \citeauthor{Wang07}, \citeyear{Wang07};
\citeauthor{Liu08}, \citeyear{Liu08}) indicating that the
overlying and surrounding magnetic field is an important factor to
determine if the flare is associated with a CME.  
Moreover, the field is largely open around the
  location of the eruptive flare (Figure~9(b)), indicating less
  strapping force.
Another example comes from 
late March 2008, when
three active regions (AR~10987, 10988, and 10989) emerged around the
same time in the same hemisphere and in a narrow longitude span of 
$\approx$60$\arcdeg$.  In terms of the sunspot area and the $R$ parameter,
AR~10988 was expected to be more active, but the two energetic CMEs
avoided it and occurred in the other regions 
(\citeauthor{Nitta11}, \citeyear{Nitta11}).  It is possible that the
presence of open field in AR~10987 and AR~10989, and the lack thereof
in AR~10988, may be an important factor.

\section{Summary} %%%%%%%%%%%%%%%%%%%%%%%%%%%%%%%%%%%%%%%%
      \label{S-Summary} 

The main objective of this article is to document the low association
of flares with CMEs during the solar minimum following Cycle 23.  
We have isolated a 34-month period that is not too close to 
the times of elevated activity in Cycle 23 or 24. 
We carefully compare EUVI and coronagraph data during the period to 
associate each flare with a CME.  Such an analysis was not
part of the EUVI flare catalog compiled by 
\citeauthor{Aschwanden09} \shortcite{Aschwanden09}, 
or the mission-long version recently generated by 
\citeauthor{Aschwanden13} \shortcite{Aschwanden13}. 
It may be relatively straightforward to automate flare detection in
high-cadence EUV images.  However, it is necessary to compare
EUV and white-light images carefully before finding pairs of associated
flares and CMEs.  Today this is still done manually.

Here we define CMEs as outflows that are 
not too narrow and are found beyond the heliocentric
distance of 5\,R$_{\sun}$, in order to distinguish them from
other processes such as waves and jets.  In a way we follow
the criticism of \citeauthor{Patsourakos09} \shortcite{Patsourakos09}
against the ``careless'' use of the term ``CME'', which may give rise to
unnecessary confusion when discussing solar eruptive phenomena.
We have found only 35 CME-associated flares, 24 of which are B-class
or less intense.  This reminds us that a CME can be launched without
strong magnetic reconnection responsible for intense flares, even
though extremely energetic CMEs tend to be associated with intense flares.
% (e.g., \citeauthor{Gopalswamy05}, \citeyear{Gopalswamy05}). 
Some active region properties, such as the sunspot area, the $R$ parameter, 
% (\citeauthor{Schrijver07}, \citeyear{Schrijver07}) 
and complexity on
the basis of the Mount Wilson sunspot classification scheme, may
largely account for intense flares.  However, CMEs may be more intimately
related to other properties, such as the co-existence of bipoles with
different polarities, weaker overlying and surrounding field, and 
proximity to open field regions. 
% the sense of magnetic helicity (\citeauthor{Welsch11}, \citeyear{Welsch11}).  
% the surrounding and overlying fields also appear to play a role in
% determining how eruptive a flare is 
% (\citeauthor{Akiyama07}, \citeyear{Akiyama07}; 
% \citeauthor{Wang07}, \citeyear{Wang07}). 

\citeauthor{Hudson01} \shortcite{Hudson01} discussed how to observe
CMEs without a coronagraph.  The proposed CME signatures in EUV images 
are often observed in the flares that we study, but none of them are observed
in all of the CME-associated flares.  We have not tested the
association in the opposite direction, i.e. how well do they 
serve as sufficient
conditions for CMEs, except for the case of EUV waves.  For EUV
waves, 13 of 34 events (38\,\%) are not associated with CMEs.  This
indicates that EUV waves 
may not be as tightly linked with CMEs as 
often thought ({\it e.g.} \citeauthor{Cliver05}, \citeyear{Cliver05}; 
\citeauthor{PFChen09}, \citeyear{PFChen09}).
This is consistent with the recent work using AIA data
\cite{Nitta13b}. 
The other three observables may also fail to be sufficient conditions
for CMEs.  In particular, we often see Type III bursts not associated
with a CME.  Therefore, we still cannot rely on low coronal
observations alone to detect and characterize CMEs.  Different
manifestations of CMEs in the low corona suggest that there may be
more than one mecahnism or a particular magnetic topology for launching CMEs.

\begin{acks}
 This work has been supported by the NASA STEREO mission under NRL
 Contract No. N00173-02-C-2035.  We thank the referee
   for useful comments that are largely reflected in the revised manuscript.
\end{acks}

\end{article} 

\end{document}